\documentclass[twocolumn,showpacs,preprintnumbers]{revtex4}
\usepackage{graphicx}
\usepackage{dcolumn}
\usepackage{bm}

\input{tcilatex}

\begin{document}

\title{Multi-object searching algorithm using subgrouped oracles}
\author{Jin-Yuan Hsieh$^{1}$ }
\author{Che-Ming Li$^{2}$}
\author{Der-San Chuu$^{2}$}
\affiliation{$^{1}$Department of Mechanical Engineering, Ming Hsin Institute of
Technology, Hsinchu 30441, Taiwan.}
\affiliation{$^{2}$Institute and Department of Electrophysics, National Chiao Tung
University, Hsinchu 30050, Taiwan.}

\begin{abstract}
We present in this work, if a set of well organized suboracles is available,
a quantum algorithm for multiobject search with certainty in an unsorted
database of $N$ items. Depending on the number of the objects, the technique
of phase tunning is included in the algorithm. If one single object is to be
searched, this algorithm performs a factor of two improvement over the best
algorithm for a classical sorted database. While if the number of the
objects is larger than one, the algorithm requires slightly less than $\log
_{4}N$ queries, but no classical counterpart exists since the resulting
state is a superposition of the marked states.
\end{abstract}

\maketitle

\address{ Tung University, Hsinchu 30050, Taiwan.}

Quantum mechanical algorithms are much more powerful than their classical
counterparts becsause they are quantum physical processes possessing the
unique feature of quantum parallelism due to superpositions and
entanglements of quantum states. The quantum parallelism potentially can
bring about an exponential speedup in running time over the classical
computation processing. A typical example is the factorizing algorithm
discovered by Shor\cite{shor} .\ The running time required in Shor's
algorithm is proportional to $O(n^{2})$ for the factorization of a number
registered in $n$ bits, while the running time for the classical algorithm
would take $O(\exp (n^{1/3}))$. This is the first quantum algorithm that
reduces a classical NP problem to a P problem, showing a quantum computer,
if it can ever be built, can transcend any classical one. One the other
hand, Grover\cite{grover1}\cite{grover2} proposed the quantum searching
algorithm, which performs a quadratic speedup over its classical
counterpart. If there is an unsorted database of $N$ items, and out of which
only one marked item satisfies a given condition, then using Grover's
algorithm one will find the object in $O(\sqrt{N})$ quantum mechanical
steps, instead of $O(N)$ classical steps. The Grover algorithm can be
extended to find $M$ objects out of a database of size $N$. It would then
required $O(\sqrt{N/M})$ iterations to search the superposition of these $M$
marked states. The superposition of the marked states is an unique feature
that only quantum algorithm can search and therefore no classical algorithms
can ever do. It has been shown that Grover$^{\prime }$s original algorithm
is optimal \cite{optimal1}\cite{optimal2}\cite{optimal3}, in the sense it
requires the minimal queries to undertake the search task. Although the
quantum searching algorithm did not provide an exponential speedup, it
indeed greatly improved in reducing the exhausting effort needed in the
classical counterpart. Up to date the searching problem remains an NP
problem.

Grover's algorithm utilizes a designed unitary operator, the Grover
operator, successively on an uniform superposition of all the possible
states to enhance the probability of the marked state. The Grover operator
is a product of two operators with specific functions, namely, the selective
inversion operator $G_{\tau }$ and the inverse-about-average operator $-G_{%
\overline{0}}$. Expressed mathematically, the Grover operator is given by $%
G=-G_{\overline{0}}$ $G_{\tau }$. Iterating $G$ about $(\pi /4)\sqrt{N}$
times on the initial state and then measuring, one will obtain the marked
state. Alternatively, the searching probelm can be rephrased in terms of the
action of an oracle. To carry out the search, one requires an oracle $f(x)$
with output $0$ or $1$ and, during the processing, is assigned to find a
specific marked item $\tau $ corresponding to the query $f(x)=1$, for $%
x=\tau $ and $f(x)=0$, for $x\neq \tau $. In the Grover algorithm each
action of $G_{\tau }$ needs one oracle call, so the quantum searching task
consume at least $(\pi /4)\sqrt{N}$ queries.

Instead of using one single oracle in Grover's algorithm for the search in
an unsorted database, Patel\cite{patel} recently suggested that if a set of
factorized oracles is available, one can locate a desired item in an
unsorted database using $O(\log _{4}N)$ queries, which is a factor of two
improvement over the best search algorithm for a classical sorted database.
Patel's scenario, in terms of the action of oracle calls, is to search a
marked state

\begin{equation}
\left| \tau \right\rangle =\left| \tau ^{(n/2)}\right\rangle \otimes
...\otimes \left| \tau ^{(2)}\right\rangle \otimes \left| \tau
^{(1)}\right\rangle
\end{equation}%
satisfying the factorizable function

\begin{equation}
f(\tau )=f_{n/2}(\tau ^{(n/2)})...f_{2}(\tau ^{(2)})f_{1}(\tau ^{(1)})=1,
\end{equation}%
provided that the state $\left| \tau \right\rangle $ is registered in $n$($%
=\log _{2}N$) qubits while each of the factorized state $\left| \tau
^{(j)}\right\rangle $, $j=1,...,n/2$, is in two qubits. The outline of this
factorized search algorithm is arranged to successively find the factorized
state $\left| \tau ^{(j)}\right\rangle $ in the corresponding
four-dimensional subspace in a single iteration according to the
corresponding factorized oracle $f_{j}(x)$, $x\in \{0,1\}^{2}$, so the total
queries are $n/2$. The factorized search algorithm is successful in
searching one single marked state since the single state is natually
factorizable. It turns out to be, however, that this algorithm will not be
applicable to a multi-solution searching problem. As mentioned, if $M$
marked states are to be searched, a quantum algorithm is designed to enhance
the probability amplitued of the superposition of these objects. The
superposition is an entangled state, so the factorized search algorithm
developed by Patel fails to solve the milti-solution searching problem. In
this work, we are thus motivated to show how to search the superposition of
the $M$ marked states in $O(\log _{4}N)$ iterations provided that a specific
set of subgrouped oracles is available.

Suppose we have an unsorted database of $N$ items at hand and according to
its size we need $n$($=\log _{2}N$) qubits to register the state in the
database. The $n$ qubits are divided into $\eta +1$ subgroups in which all
are of two qubits except one $n_{0}$-qubit subgroup, so $n=2\eta +n_{0}$,
where $n_{0}$ is chosen depending on the number of the desired items $M$ and
will be defined in what follows. The marked states then can be expressed as

\begin{eqnarray}
\left| \tau _{j}\right\rangle &=&\left| \tau _{j}^{(\eta +1)}...\tau
_{j}^{(2)}\tau _{j}^{(1)}\right\rangle  \nonumber \\
&=&\left| \tau _{j}^{(\eta +1)}\right\rangle \otimes ...\otimes \left| \tau
_{j}^{(2)}\right\rangle \otimes \left| \tau _{j}^{(1)}\right\rangle ,\text{ }%
j=1,...,M,
\end{eqnarray}%
where

\begin{equation}
\tau _{j}^{(1)}\in \{1,0\}^{n_{0}},\text{ }n_{0}=\left\lfloor \log
_{2}(4M)\right\rfloor \geq 2,
\end{equation}

\begin{equation}
\tau _{j}^{(k)}\in \{1,0\}^{2},\text{ }k=2,\text{ }...,\text{ }\eta +1.
\end{equation}%
The ''floor'' symbol $\lfloor $ $\rfloor $ in (4) denotes the largest
integer smaller than the value in it. Although the ordering of the $n_{0}$%
-qubit factorized states can be arbitrary specified, we have designated them
as $\tau _{j}^{(1)}$ in the first subspace for convenience. As mentioned,
our purpose is to outlet the superposition of the marked states, namely,

\begin{equation}
\left| \tau \right\rangle =\frac{1}{\sqrt{M}}\sum_{j=1}^{M}\left| \tau
_{j}\right\rangle =\frac{1}{\sqrt{M}}\sum_{j=1}^{M}\left| \tau _{j}^{(\eta
+1)}...\tau _{j}^{(2)}\tau _{j}^{(1)}\right\rangle .
\end{equation}
The initial state is prepared in the uniform superposition of all the
possible states, given by

\begin{equation}
\left| s\right\rangle =W_{n}\left| \overline{0}\right\rangle =\frac{1}{\sqrt{%
N}}\sum_{y\in \{0,1\}^{n}}\left| y\right\rangle
\end{equation}%
where $W_{n}$ is the $n$-qubit Walsh-Hadamard transformation and $\left| 
\overline{0}\right\rangle $ is the standard state with $0$ in every qubit.

By expression (3) we realize that the global, $2^{n}$-dimensional Hilbert
space $\mathcal{H}$ for the search is the tensor product of $\eta +1$
subspaces, expressed as $\mathcal{H}=\mathcal{H}_{2}^{\otimes \eta }\otimes 
\mathcal{H}_{n_{0}}$, where the subspace $\mathcal{H}_{n_{0}}$ is $2^{n_{0}}$%
-dimensional and the subspace $\mathcal{H}_{2}$ is four-dimensional. The
scenario of the algorithm contains $\eta +1$ sequential stages. It begins
with finding the $M$ factorized states $\left| \tau _{j}^{(1)}\right\rangle $
in the subspace $\mathcal{H}_{n_{0}}$ in just one searching step. Knowing
that Grover's algorithm promises to search $M$ objects with certainty in a
database containing $4M$ items by applying the Grover operator once, we
therefore choose $n_{0}=\log _{2}(4M)$ for the first subgrouped qubits. Even
if $\log _{2}(4M)$ is not an integer, we can also search the objects with
certainty in just one searching step by choosing $n_{0}=\left\lfloor \log
_{2}(4M)\right\rfloor $ and using the technique of phase tunning\cite{hsieh
and li}. In the $2^{n_{0}}$-dimensional subspace $\mathcal{H}_{n_{0}}$, we
apply the subgrouped Grover operator $G_{1}$, given by 
\begin{equation}
G_{1}=-(I+(e^{i\phi }-1)\left| s^{(1)}\right\rangle \left\langle
s^{(1)}\right| )(I+(e^{i\phi }-1)\left| \tau ^{(1)}\right\rangle
\left\langle \tau ^{(1)}\right| ),
\end{equation}%
where

\begin{equation}
\left| \tau ^{(1)}\right\rangle =\frac{1}{\sqrt{M}}\sum_{j=1}^{M}\left| \tau
_{j}^{(1)}\right\rangle ,\text{ }\left| s^{(1)}\right\rangle =\frac{1}{\sqrt{%
2^{n_{0}}}}\sum_{y\in \{0,1\}^{n_{0}}}\left| y\right\rangle ,
\end{equation}%
and the phase angle, according to Hsieh and Li\cite{hsieh and li}, is given
by 
\begin{equation}
\phi =\phi (M)=2\sin ^{-1}(\sqrt{\frac{2^{n_{0}}}{4M}}).
\end{equation}%
Notice that if $\log _{2}(4M)$ is an integer, $\phi =\pi $ and $G_{1}$
reduces to the original Grover operator defined in $\mathcal{H}_{n_{0}}$,
while if $\log _{2}(4M)$ is not an integer, the choice $n_{0}=\left\lfloor
\log _{2}(4M)\right\rfloor $ leads to the smallest deviation of the phase $%
\phi $ form $\pi $. The oracle used in the first subgrouped Grover operator $%
G_{1}$ should be

\begin{eqnarray}
f_{1}(y) &=&1,\text{ \ \ \ }y=\tau _{j}^{(1)},\text{ \ }j=1,...,M, \\
f_{1}(y) &=&0,\text{ \ \ \ }y\neq \tau _{j}^{(1)},\text{ \ }j=1,...,M. 
\nonumber
\end{eqnarray}

In the first stage the searching operator that we apply on the initial state 
$\left| s\right\rangle $ is

\begin{equation}
S_{1}=I_{2}^{\otimes \eta }\otimes G_{1},
\end{equation}%
where $I_{2}$ denotes the identity operator in the subspace $\mathcal{H}_{2}$%
. The outlet at the end of this stage then becomes

\begin{equation}
\left| t_{1}\right\rangle =S_{1}\left| s\right\rangle =(\frac{1}{\sqrt{%
2^{n-n_{0}}}}\sum_{y\in \{0,1\}^{n-n_{0}}}\left| y\right\rangle )\otimes
\left| \tau ^{(1)}\right\rangle .
\end{equation}%
The outlet state $\left| t_{1}\right\rangle $ can be rewritten, for
convenience in the depiction of the next stage,

\begin{equation}
\left| t_{1}\right\rangle =(\frac{1}{\sqrt{2^{n-n_{0-2}}}}\sum_{y\in
\{0,1\}^{n-n_{0}-2}}\left| y\right\rangle )\otimes \left|
s^{(2)}\right\rangle ,
\end{equation}%
where

\begin{equation}
\left| s^{(2)}\right\rangle =\frac{1}{\sqrt{4M}}\sum_{j=1}^{M}(\sum_{x\in
\{0,1\}^{2}}\left| x\right\rangle )\otimes \left| \tau
_{j}^{(1)}\right\rangle
\end{equation}%
is located in the subspace $\mathcal{H}_{2}\otimes \mathcal{H}_{n_{0}}$.
Now, observing that in the subspace $\mathcal{H}_{2}\otimes \mathcal{H}%
_{n_{0}}$ the state $\left| s^{(2)}\right\rangle $ is a superposition of $4M$
nonvanishing orthonormal states $\left| x\tau _{j}^{(1)}\right\rangle $, $%
x\in \{0,1\}^{2}$, $j=1,...,M$, and in them the $M$ factorized marked states 
$\left| \tau _{j}^{(2)}\tau _{j}^{(1)}\right\rangle $ are embedded. We then
once again can apply a second subgrouped Grover operator $G_{2}$ to search
these factorized marked states with certainty in just one searching step.
The second subgrouped Grover operator $G_{2}$ is defined in the subspace $%
\mathcal{H}_{2}\otimes \mathcal{H}_{n_{0}}$, given by

\begin{equation}
G_{2}=-(I-2\left| s^{(2)}\right\rangle \left\langle s^{(2)}\right|
)(I-2\left| \tau ^{(2)}\right\rangle \left\langle \tau ^{(2)}\right| ),
\end{equation}%
where

\begin{equation}
\left| \tau ^{(2)}\right\rangle =\frac{1}{\sqrt{M}}\sum_{j=1}^{M}\left| \tau
_{j}^{(2)}\tau _{j}^{(1)}\right\rangle .
\end{equation}%
The oracle implemented in $G_{2}$ is of course defined by

\begin{eqnarray}
f_{2}(y) &=&1,\text{ \ \ \ }y=\tau _{j}^{(2)}\tau _{j}^{(1)},\text{ \ }%
j=1,...,M, \\
f_{2}(y) &=&0,\text{ \ \ \ }y\neq \tau _{j}^{(2)}\tau _{j}^{(1)},\text{ \ }%
j=1,...,M.  \nonumber
\end{eqnarray}%
The searching operator that we apply in the second stage then is

\begin{equation}
S_{2}=I_{2}^{\otimes (\eta -1)}\otimes G_{2},
\end{equation}%
and the outlet at the end of the second stage is

\begin{eqnarray}
\left| t_{2}\right\rangle &=&S_{2}\left| t_{1}\right\rangle
=S_{2}S_{1}\left| s\right\rangle \\
&=&(\frac{1}{\sqrt{2^{n-n_{0-2}}}}\sum_{y\in \{0,1\}^{n-n_{0}-2}}\left|
y\right\rangle )\otimes \left| \tau ^{(2)}\right\rangle  \nonumber \\
&=&(\frac{1}{\sqrt{2^{n-n_{0-4}}}}\sum_{y\in \{0,1\}^{n-n_{0}-4}}\left|
y\right\rangle )\otimes \left| s^{(3)}\right\rangle ,  \nonumber
\end{eqnarray}%
where the state $\left| s^{(3)}\right\rangle $ is written

\begin{equation}
\left| s^{(3)}\right\rangle =\frac{1}{\sqrt{4M}}\sum_{j=1}^{M}(\sum_{y\in
\{0,1\}^{2}}\left| x\right\rangle )\otimes \left| \tau _{j}^{(2)}\tau
_{j}^{(1)}\right\rangle ,
\end{equation}%
which locates in the third subspace $\mathcal{H}_{2}^{\otimes 2}\otimes 
\mathcal{H}_{n_{0}}$ and is also a superposition of the $4M$ nonvanishing
orthonormal states $\left| x\tau _{j}^{(2)}\tau _{j}^{(1)}\right\rangle $, $%
x\in \{0,1\}^{2}$, $j=1,...,M$, with the $M$ factorized marked states $%
\left| \tau _{j}^{(3)}\tau _{j}^{(2)}\tau _{j}^{(1)}\right\rangle $ embedded
in them. So the scenario obvisouly can proceed staightfoward to the final
statge following the same way as depicted from the first to the second
stage. That is, in a next stage we add two more qubits to the former
subgrouped qubits, so a next subspace is the one enlarged by multiplying the
dimensions of the former subspace by four, and apply a searching operator
like those given by (12) and (19) and a corresponding subgrouped oracle like
those expressed in (11) and (18).

As a result, we can thus derive the general formula for the present
searching algorithm. Excluding the first stage, where as we can apply
expressions (8)-(13), in the $k$-th stage, $2\leq k\leq \eta +1$, the
searching operator is

\begin{equation}
S_{k}=I_{2}^{\otimes (\eta -k+1)}\otimes G_{k},
\end{equation}%
and the $k$-th subgrouped Grover operator $G_{k}$ is defined in the subspace 
$\mathcal{H}_{2}^{\otimes (k-1)}\otimes \mathcal{H}_{n_{0}}$, given by%
\begin{equation}
G_{k}=-(I-2\left| s^{(k)}\right\rangle \left\langle s^{(k)}\right|
)(I-2\left| \tau ^{(k)}\right\rangle \left\langle \tau ^{(k)}\right| ),
\end{equation}%
where

\begin{equation}
\left| \tau ^{(k)}\right\rangle =\frac{1}{\sqrt{M}}\sum_{j=1}^{M}\left| \tau
_{j}^{(k)}...\tau _{j}^{(2)}\tau _{j}^{(1)}\right\rangle ,
\end{equation}%
and

\begin{equation}
\left| s^{(k)}\right\rangle =\frac{1}{\sqrt{4M}}\sum_{j=1}^{M}(\sum_{y\in
\{0,1\}^{2}}\left| x\right\rangle )\otimes \left| \tau _{j}^{(k-1)}...\tau
_{j}^{(2)}\tau _{j}^{(1)}\right\rangle .
\end{equation}

The corresponding subgrouped oracle required in the $k$-th stage is defined
by

\begin{eqnarray}
f_{k}(y) &=&1,\text{ \ \ \ }y=\tau _{j}^{(k)}...\tau _{j}^{(2)}\tau
_{j}^{(1)},\text{ \ }j=1,...,M, \\
f_{k}(y) &=&0,\text{ \ \ \ }y\neq \tau _{j}^{(k)}...\tau _{j}^{(2)}\tau
_{j}^{(1)},\text{ \ }j=1,...,M.  \nonumber
\end{eqnarray}%
The outlet state at the end of the $k$-th stage then is

\begin{eqnarray}
\left| t_{k}\right\rangle &=&S_{k}\left| t_{k-1}\right\rangle
=S_{k}...S_{2}S_{1}\left| s\right\rangle \\
&=&(\frac{1}{\sqrt{2^{n-n_{0-}2(k-1)}}}\sum_{y\in
\{0,1\}^{n-n_{0}-2(k-1)}}\left| y\right\rangle )\otimes \left| \tau
^{(k)}\right\rangle .  \nonumber
\end{eqnarray}%
As $k=\eta +1$, i.e., as we arrive at the final stage, we will reach the
final state at the end,

\begin{eqnarray}
\left| t_{\eta +1}\right\rangle &=&S_{\eta +1}\left| t_{\eta }\right\rangle
=S_{\eta +1}...S_{2}S_{1}\left| s\right\rangle \\
&=&\left| \tau ^{(\eta +1)}\right\rangle =\frac{1}{\sqrt{M}}%
\sum_{j=1}^{M}\left| \tau _{j}^{(\eta +1)}...\tau _{j}^{(2)}\tau
_{j}^{(1)}\right\rangle  \nonumber \\
&=&\left| \tau \right\rangle ,  \nonumber
\end{eqnarray}%
which is exactly the object that we search for. It should be mentioned that
a projected / measurement operator, although not shown, has been combined
within the searching operator used in each of the stages to make sure the
eliminated states do not take part in the subsequent stages of the
algorithm. To summarize, totally $\eta +1$ queries are required in our
subgrouped searching algorithm since in each stage of the scenario only one
query is utilized. The number $\eta +1$, being equal to $(n-n_{0}+2)/2$, is
in fact smaller than $n/2$ for $M>1$. As $M=1$, our algorithm will reduce to
the factorized algorithm presented by Patel. A circuit representation of the
present algorithm is shown in Fig. 1 to schematically depict the whole
scenario.

\bigskip \FRAME{ftbphFU}{3.2422in}{2.5598in}{0pt}{\Qcb{Circuit for the
subgrouped searching algorithm. The subgrouped Grover operator $G_{k}$ in
the $k$-th stage is applied in a subspace of $n_{0}+2(k-1)$ qubits. $W_{n}$
and $W_{\protect\eta +1}$ are the $n$-qubit and $(\protect\eta +1)$-qubit
Walsh-Hadamard transformations, respectively.}}{}{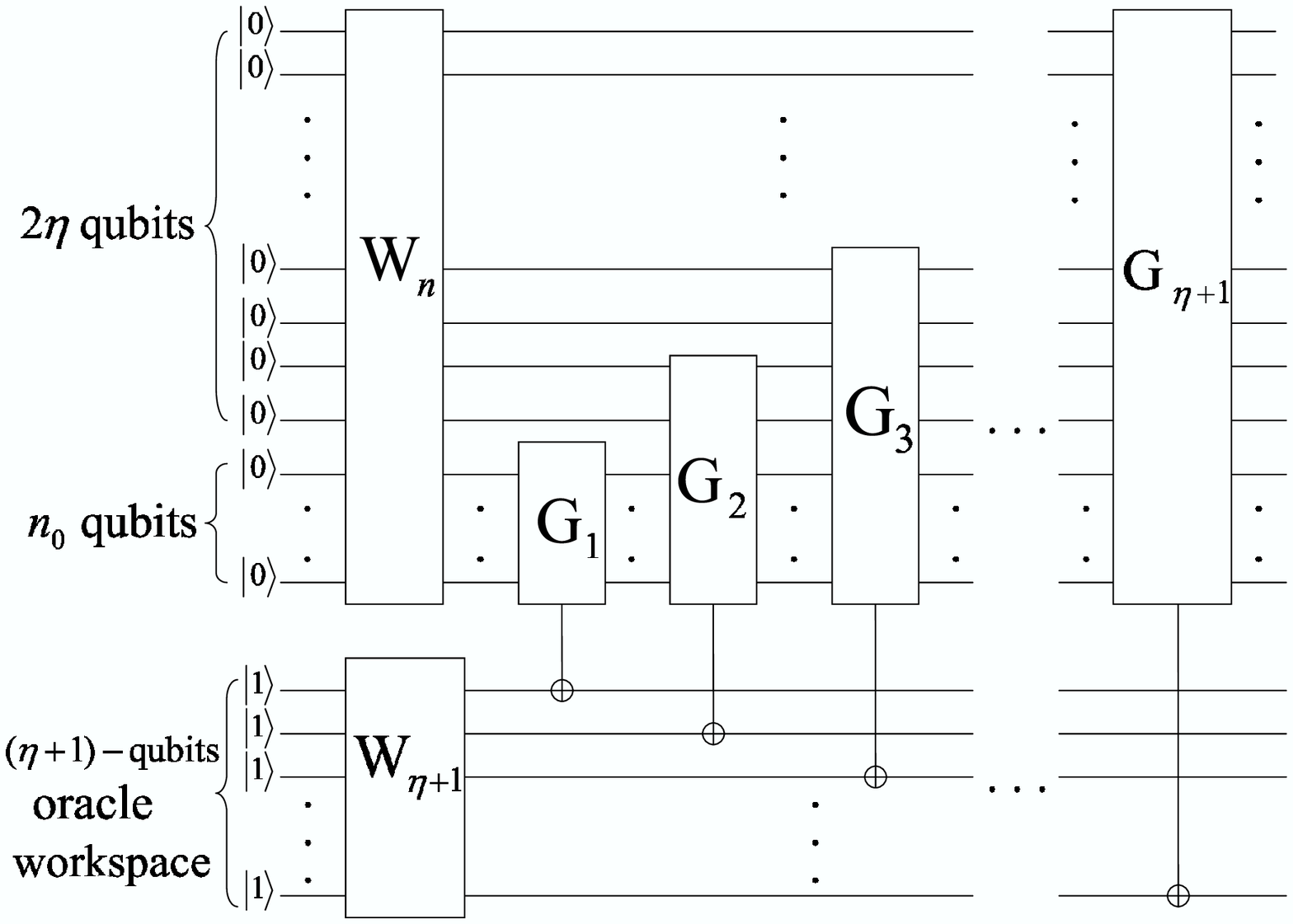}{\special%
{language "Scientific Word";type "GRAPHIC";maintain-aspect-ratio
TRUE;display "USEDEF";valid_file "F";width 3.2422in;height 2.5598in;depth
0pt;original-width 7.8871in;original-height 8.2382in;cropleft
"0.0643842";croptop "1.0335586";cropright "1.0297842";cropbottom
"0.3058586";filename 'fig1.eps';file-properties "XNPEU";}}

In this work, we have shown a subgrouped algorithm using only $O(\log _{4}N)$
queries in searching $M$ objects out of an unsorted database of $N$ items,
if in advance a set of subgrouped oracles is available. The subgrouped
oracles required in the algorithm are expressed in (11) and (26), while the
searching operator are defined by (12) and (22), with the aides of (8)-(10)
and (23)-(25), respectively. The first searching operator $S_{1}$ is more
particular than the others because in it the technique of phase tunning is
necessary for the action of the subgrouped Grover operator $G_{1}$
especially when $\log _{2}(4M)$ is not an integer. Using this subgrouped
searching algorithm, one will require $O(\log _{4}N)$ queries in the search
of multi-object. The resulting state at the end of the algorithm is the
superposition of the $M$ marked states, which is an unique feature of the
quantum algorithm and can not be performed by any classical algorithm.
Difficulties, however, may arise from how to physically implement the black
boxes associated with the subgrouped oracles, but these are not of concern
in the complexity analysis of this work and should be studied in a furture
work.

\end{document}